\documentclass[twocolumn,A4]{article}
\usepackage[dvips]{graphics}
\begin{document}
\onecolumn
\begin{center}
{\bf{\Large Strange behavior of persistent currents in small
Hubbard rings}}\\
~\\
Santanu K. Maiti, J. Chowdhury and S. N. Karmakar$^*$\\
~\\
{\em Saha Institute of Nuclear Physics, 1/AF, Bidhannagar,
Kolkata 700 064, India}\\
~\\
~\\
~\\
{\bf Abstract}
\end{center}

We show exactly that small Hubbard rings exhibit unusual kink-like
structures giving anomalous oscillations in persistent current.
Singular behavior of persistent current disappears in some cases.
In half-filled systems mobility gradually drops to zero with interaction,
while it converges to some finite value in non-half-filled cases.
\vskip 2cm
\begin{flushleft}
\end{flushleft}
\vskip 3in
\noindent
{\bf ~$^*$Corresponding Author}:

Email address:  sachin@cmp.saha.ernet.in
\newpage
\twocolumn
Since the pioneering work of B\"{u}ttiker, Imry, and Landauer~\cite{butt},
the phenomenon of persistent current in mesoscopic systems has been 
discussed theoretically in various papers ~\cite{land,cheu1,mont,
alts,von,schm,yu,abra,bouz,giam,pichard,burme}. 
Although this phenomenon is thought to be
qualitatively understood within the framework 
of one-electron theory ~\cite{butt,land,cheu1,mont, alts,von}, 
it has proved difficult to explain recent
experimental results~\cite{levy,maily,jari,deb}.
The unexpected large amplitude of the measured
currents, the observation of currents with both $hc/e$ and $hc/2 e$ flux 
periodicity, and, the sign of the currents are the major subjects of
disagreement. There are many theoretical attempts to explain the 
experimental results, and it is generally believed 
that disorder~\cite{alts,von,abra,bouz,giam} and
electron-electron correlation~\cite{schm,yu,abra,bouz,giam,pichard,burme}
 are the key factors which can resolve
the controversy between theory and experiment. However, no consensus on the
role of interaction and disorder has yet been reached. The explanation
of the experimental results become even much more illusive since 
recently Kravtsov {\em et al.}~\cite{krav} 
have shown that additional currents
may be generated in the rings by other mechanisms which are not
experimentally distinguishable from persistent current. 

To understand the precise role of correlation, in this letter 
we concentrate on the exact calculation of persistent
current and Drude weight in ordered $1$D Hubbard rings threaded by
magnetic flux.
Our study reveals that certain aspects of the many-body effects on 
persistent current have not been discussed in the literature, and
we will see that Hubbard correlation leads to many remarkable effects.
We limited ourselves to small Hubbard rings and
these studies might be also helpful to
understand the physical properties of TTF-TCMQ conductors, various
aromatic molecules and systems of connected quantum dots~\cite{yu}. 
With the new technological advancements, it is now possible to
fabricate such small rings, and, in a recent experiment Keyser
{\em et al.} \cite{keyser} reported the evidence of anomalous
Aharonov-Bohm oscillations from the transport measurements on small
quantum rings with less than ten electrons. In these small rings
with few number of electrons, the electron-electron interaction
becomes important as the Coulomb potential is not screened much, and we will
show that the electron-electron interaction leads to similar anomalous
oscillations in the persistent current as a function of the magnetic
flux.

We use Hubbard model to represent the system, which for a $N$-site ring
enclosing a magnetic flux $\phi$ (in units of elementary flux quantum
$\phi_0=h c/e$) can be expressed as
\begin{equation}
H=t\sum_{i,\sigma}\left[e^{i \Theta} c_{i,\sigma}^
{\dagger} c_{i+1,\sigma}+ h.c. \right]
+U \sum_{i}n_{i\uparrow}n_{i\downarrow} 
\label{hamil}
\end{equation}
where $\Theta =2 \pi \phi/N $, and the other symbols have their
usual meanings. We take $t=-1$ and use the units $c=1$, $e=1$.

At zero temperature, the persistent current in the ring threaded by
flux $\phi$ is determined by~\cite{cheu1}
\begin{equation}
I(\phi) = - \frac{\partial{E_{0}(\phi)}}{\partial{\phi}} ,
\label{curr}
\end{equation}
where $E_{0}(\phi)$ is the ground state energy. We evaluate this
quantity exactly to understand unambiguously the role of 
interaction on persistent current, and this is achieved by
exact diagonalization of the many-body Hamiltonian Eq.~(\ref{hamil}). 

We also consider the response of the system to a uniform 
time-dependent electric field in terms of the Drude weight 
~\cite{scal1,scal2} $D$,
a closely related parameter that characterizes the conducting nature
of the system as originally noted by Kohn~\cite{kohn}. 
The Drude weight can be calculated from the relation~\cite{bouz} 
\begin{equation}
D = \frac{N}{4\pi^2}\left(\frac{\partial{^2E(\phi)}}{\partial{\phi}^{2}}
\right)_{\phi = \phi_{min}} ,
\label{drude}
\end{equation}
where $\phi_{min}$ is the location of the minimum of $E_{0}(\phi)$. 
Now we present below the results of our calculations for the
various systems.

\begin{table*}
\caption{Eigenvalues ($\lambda$) and eigenstates
of two opposite spin electrons for N=3.}
\label{table}
\begin{tabular}{|c|c|c|c|c|c|c|}  
\hline
Total & \multicolumn{3}{c|}{$U=0$} & \multicolumn{3}{c|}{$U=2$} \\ \cline{2-7}
spin $S$ & $\lambda$ & Degeneracy & Eigenstate & $\lambda$ & Degeneracy & 
Eigenstate \\ \hline
 & $-4$ & 1 & $(-\frac{1}{\sqrt{2}}, -\frac{1}{\sqrt{2}}, -\frac{1}{\sqrt{2}},
-1, -1,1)$ & $-2 \sqrt{3}$ & 1 & $(a, a, a, -1, -1, 1)$
\\ \cline{2-7} 
 &$-1$ & 2 & $(0, \sqrt{2}, -\sqrt{2}, 1, 0, 1)$ & 0 & 2 & $(0, \frac{1}{\sqrt{2
}}, -\frac{1}{\sqrt{2}}, 1, 0, 1)$ \\
 &  &  & $(-\sqrt{2}, 0, \sqrt{2}, -1, 1, 0)$ &  &  & $(-\frac{1}{\sqrt{2}},
0, \frac{1}{\sqrt{2}}, -1, 1, 0)$ \\ \cline{2-7} 
0 & 2 & 3 & $(\frac{1}{\sqrt{2}}, 0, \frac{1}{\sqrt{2}}, 0, 0, 1)$ & 3 &
2 & $(0, -\sqrt{2}, \sqrt{2}, 1, 0, 1)$ \\ 
 &  & & $(0, -\frac{1}{\sqrt{2}}, -\frac{1}{\sqrt{2}}, 0, 1, 0)$ & & &
 $(\sqrt{2}, 0, -\sqrt{2}, -1, 1, 0)$ \\
 & & & $(-\frac{1}{\sqrt{2}}, -\frac{1}{\sqrt{2}}, 0, 1, 0, 0)$ & $2 \sqrt{3}$
& 1 & $(b, b, b, -1, -1, 1)$ \\ \cline{1-7}
 & $-1$ & 2 & $(1, 0, 1)$ & $-1$ & 2 & $(1, 0, 1)$ \\
1 & & & $(-1, 1, 0)$ & & & $(-1, 1, 0)$ \\
 & 2 & 1 & $(-1, -1, 1)$ & 2 & 1 & $(-1, -1, 1)$ \\ \hline
\end{tabular}
\end{table*}

We first consider two electron systems, and to have
a deeper insight to the problem, we begin our
discussion with the simplest possible system which can be treated
analytically up to certain level. This is the case of a three-site
ring consisting of two opposite spin electrons. The Hamiltonian is a
$(9\times 9)$ matrix which can be block diagonalized to $(6\times 6)$
and $(3\times 3)$ matrices by proper choice of the basis states. This
can be done by constructing basis states for each sub-space with a
definite value of the total spin $S$.
The basis set for the six-dimensional sub-space is
chosen as $\mathcal{A}\equiv$ $\{|\uparrow\downarrow,0,0>$,
$|0,\uparrow\downarrow,0>$,
$|0,0,\uparrow\downarrow>$, $(|\uparrow,\downarrow,0>-|\downarrow,
\uparrow,0>)/\sqrt{2}$, $(|0,\uparrow,\downarrow>-|0,\downarrow,\uparrow>)/
\sqrt{2}$, $(|\downarrow,0,\uparrow>-|\uparrow,0,\downarrow>)/\sqrt{2}\}$
(each with $S=0$), while that for the other sub-space is taken as
$\mathcal{B}\equiv$ $\{(|\uparrow,\downarrow,0>+|\downarrow,
\uparrow,0>)/\sqrt{2}$, $(|0,\uparrow,\downarrow>+|0,\downarrow,\uparrow>)/
\sqrt{2}$, $(|\uparrow,0,\downarrow>+|\downarrow,0,\uparrow>)/\sqrt{2}\}$
(each with $S=1$).
For $U=0$ and $U=2$, in Table~\ref{table} we list
the energy eigenvalues, eigenstates and the degeneracy of the levels of
this system in the absence of any magnetic field.
In this table we set
$a=-\sqrt{2}/(1+\sqrt{3})$ and $b=\sqrt{2}/(-1+\sqrt{3})$.
Both for $U=0$ and $U\ne0$ cases,
the two-fold degeneracy of the energy levels corresponds to the $C_{3v}$ 
symmetry of the system, and, this degeneracy gets lifted by the
magnetic flux. It is apparent from this table that the eigenvalues,
eigenstates and also the degeneracy of the levels are not affected
by correlation in the three-dimensional sub-space (the reason is that 
basis set $\mathcal{B}$ does not involve any doubly occupied site), 
but this is not true in the other sub-space.
\begin{figure}[h]
{\centering \resizebox*{8.0cm}{5.5cm}{\includegraphics{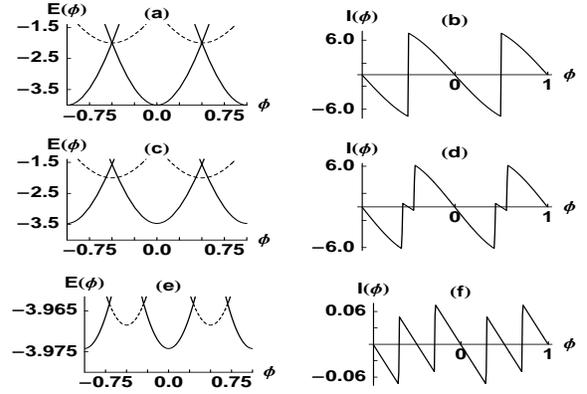}}\par}
\caption{The $E-\phi$ and $I-\phi$ curves for following two ($\uparrow,
\downarrow$) electron systems : i) $N=3$, $U=0$ in (a) \& (b); ii)
$N=3$, $U=2$ in (c) \& (d); and iii) $N=25$, $U=3$ in (e) \& (f).}
\label{energy1}
\end{figure}

The insertion of magnetic flux $\phi$ does not alter the above structure
of the Hamiltonian, though it becomes now field dependent.
In Fig.~\ref{energy1}(a) we have plotted some of the
low-lying energy levels $E(\phi)$'s
as functions of $\phi$ (say, $E-\phi$ curves) of this three-site
ring with $U=0$, and, the corresponding $I(\phi)$ versus $\phi$ curve
(say, $I-\phi$ curve) is shown in Fig.~\ref{energy1}(b).
The $E-\phi$ and $I-\phi$
curves for $U=2$ case are respectively given in Fig.~\ref{energy1}(c)
and Fig.~\ref{energy1}(d). 
In Fig.~\ref{energy1}(a) and Fig.~\ref{energy1}(c),
the dotted curves represent the $U$-independent
energy levels as mentioned above. Quite interestingly Fig.~\ref{energy1}(c)
shows that even in the presence of interaction, 
one of the $U$-independent energy levels becomes the ground state
energy level in certain intervals of $\phi$ (e.g., around $\phi=\pm 0.5$).
In other
regions of $\phi$ we have the expected results, i.e., the ground state
energy increases due to correlation and the $E_{0}-\phi$ curves become
much shallow. As a result, the usual sawtooth shape of the $I-\phi$ 
characteristics of $U=0$ case changes drastically as shown in
Fig.~\ref{energy1}(d), and the role of correlation is quite complex
rather than a simple suppression of the persistent current as predicted
earlier. A sudden change in the direction and magnitude of 
persistent current occurs, solely due to correlation, around $\phi=\pm 0.5$,
and, it forms kink-like structures in the persistent current as illustrated
in Fig.~\ref{energy1}(d). The kinks become wider
as we increase $U$, and, the most 
surprising result is that the persistent current remain invariant inside
the kinks irrespective of correlation. It is also seen that 
correlation does not affect the $\phi_{0}$ flux periodicity of the persistent
current. 

With the above background, we next investigate
the role of correlation on persistent current in larger mesoscopic rings. 
In Fig.~\ref{energy1}(e) and Fig.~\ref{energy1}(f) we
display respectively the $E-\phi$ and $I-\phi$ curves for a ring
containing two opposite spin electrons with $N=25$ and $U=2$.
These figures respectively resembles Fig.~\ref{energy1}(c)
and Fig.~\ref{energy1}(d),
and it implies that correlation plays the same
role in both the cases. Thus we can precisely say that away from 
half-filling, a ring consisting of two opposite spin electrons always
exhibits kinks in the persistent current for any non-zero value of $U$.

\begin{figure}[h]
{\centering \resizebox*{8.0cm}{5.5cm}{\includegraphics{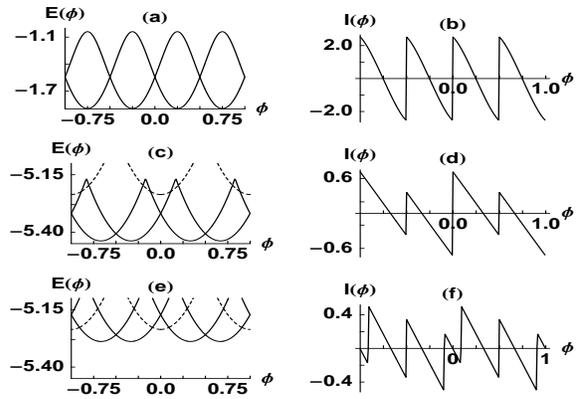}}\par}
\caption{The $E-\phi$ and $I-\phi$ curves for following three ($\uparrow,
\uparrow,\downarrow$) electron systems : i) $N=3$, $U=3$ in (a) \& (b); ii)
$N=10$, $U=3$ in (c) \& (d); and iii) $N=10$, $U=15$ in (e) \& (f).}
\label{energy2}
\end{figure}

Next we consider rings with two up and one down spin electrons
as representative examples of three electron systems. 
In Fig.~\ref{energy2}(a) and Fig.~\ref{energy2}(b), we 
respectively plot the $E-\phi$ and $I-\phi$ curves for the half-filled
situation ($N=3$ and $n=3$, where $n$ denotes the number of electrons)
with $U=3$. We find that
correlation just diminishes the magnitude of the current
compared to that of the noninteracting case. 
In Fig.~\ref{energy2}(c) and Fig.~\ref{energy2}(d) we 
respectively plot the $E-\phi$ and $I-\phi$
curves of a non-half-filled situation with $N=10$ and $U=3$,
while Fig.~\ref{energy2}(e) and Fig.~\ref{energy2}(f) are these curves
for the $U=15$ case. The behavior of the persistent
current, as a function of flux, are quite different
at low and high values of $U$. It is evident from Fig.~\ref{energy2}(c)
that for low $U$, all the $U$-independent energy
levels (dotted curves) always lie above the ground state energy
of the system, and Fig.~\ref{energy2}(d) shows that,
apart from a reduction of the
persistent current, the $I-\phi$ curve looks very similar to that of a
ring without any interaction.
On the other hands, we see from Fig.~\ref{energy2}(e) 
that for high $U$, one of the $U$-independent energy levels becomes
the ground state energy in certain intervals of $\phi$
(e.g., around $\phi=0$), and, this produces kinks in the $I-\phi$
characteristics as depicted in Fig.~\ref{energy2}(f).
Certainly there exists a critical 
value $U_c$ of correlation above which kinks appear in the $I-\phi$
curve. We have verified that these features of the energy spectra and
persistent current are the characteristics of any non-half-filled
ring with two up and one down spin interacting electrons. We also
note that the half-filled cases exhibit $\phi_{0}/2$ periodic currents, 
while the non-half-filled cases have $\phi_{0}$ periodicity.

\begin{figure}[h]
{\centering \resizebox*{8.0cm}{5.5cm}{\includegraphics{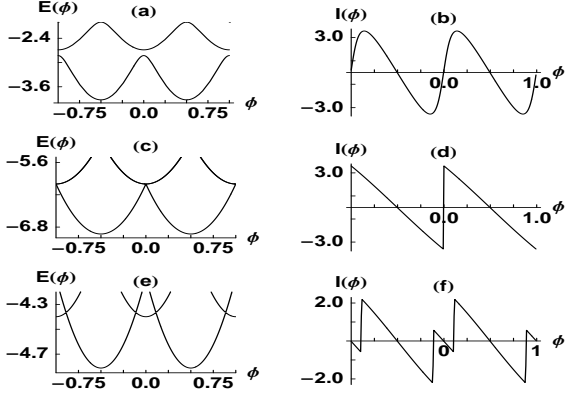}}\par}
\caption{The $E-\phi$ and $I-\phi$ curves for following four ($\uparrow,
\uparrow,\downarrow,\downarrow$) electron systems :
i) $N=4$, $U=2$ in (a) \& (b); ii)
$N=6$, $U=0$ in (c) \& (d); and iii) $N=6$, $U=6$ in (e) \& (f).}
\label{energy3}
\end{figure}

Now we consider rings with two up and two down spin electrons
as illustrative examples of four electron systems.
Let us first study the half-filled case ($N=4$ and $n=4$).
In the absence of any interaction, this system exhibits sharp discontinuity
in persistent current at certain values of $\phi$.
However, the influence of correlation is quite dramatic
and it makes $I(\phi)$ a continuous function of $\phi$ as
depicted in Fig.~\ref{energy3}(b).
This is a quite fascinating result since correlation
drastically changes the analytic behavior of $I(\phi)$, and we will see that
this result also holds true for other half-filled systems with even
number of electrons.
Away from half-filling, we study an interesting typical case of a six-site
ring with two up and two down spin electrons.
It may be considered as a
doubly ionized benzene-like ring, a system with the promise of experimental
verification of our predictions. In Fig.~\ref{energy3}(c)
and Fig.~\ref{energy3}(d) we respectively plot 
the $E-\phi$ and $I-\phi$ curves with $U=0$, while Fig.~\ref{energy3}(e)
and Fig.~\ref{energy3}(f) are these plots with $U=6$.
Fig.~\ref{energy3}(f) shows that kinks appear in
the $I-\phi$ curve (e.g., around $\phi=0$) for any non-zero value
of $U$, but we interestingly note that the kinks
are now due to $U$-dependent eigenstates. 
Both the half-filled and non-half-filled rings exhibit $\phi_{0}$ periodic
currents.

\begin{figure}[h]
{\centering \resizebox*{8.0cm}{5.5cm}{\includegraphics{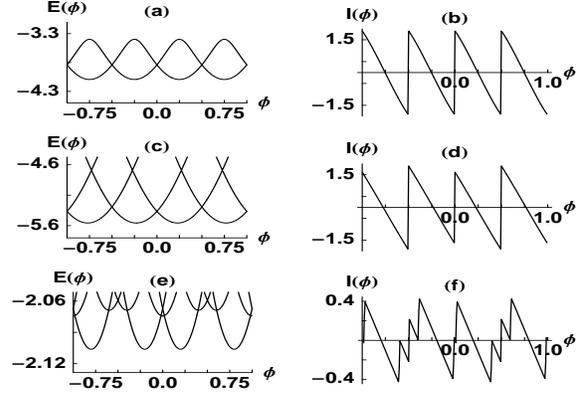}}\par}
\caption{The $E-\phi$ and $I-\phi$ curves for following five ($\uparrow,
\uparrow,\uparrow$, $\downarrow,\downarrow$) electron systems :
i) $N=5$, $U=2$ in (a) \& (b); ii)
$N=6$, $U=2$ in (c) \& (d); and iii) $N=6$, $U=120$ in (e) \& (f).}
\label{energy4}
\end{figure}

As specific cases of five electron systems, 
we study rings with three up and two down spin electrons.
The $E-\phi$ and $I-\phi$ curves for the half-filled
ring ($N=5$ and $n=5$) are plotted in Fig.~\ref{energy4}(a)
and Fig.~\ref{energy4}(b) respectively.
Interaction just diminishes the magnitude of the current 
and we find $\phi_{0}/2$ periodicity.
As a non-half-filled five electron case, we take a singly ionized 
benzene-like ring ($N=6$, $n=5$) and calculate persistent current for low and 
high values of $U$. In Fig.~\ref{energy4}(c) and
Fig.~\ref{energy4}(d) we respectively draw the $E-\phi$ 
and $I-\phi$ curves taking $U=2$, while Fig.~\ref{energy4}(e)
 and Fig.~\ref{energy4}(f) are these diagrams with $U=120$.
It is clear from Fig.~\ref{energy4}(d) and Fig.~\ref{energy4}(f)
that like the three
electron cases, kinks appear in the persistent current only after a
critical value of $U$. This reveals that the characteristic features
of persistent current in mesoscopic Hubbard rings with odd number of
electrons are almost identical. 

\begin{figure}[h]
{\centering \resizebox*{8.0cm}{4.5cm}{\includegraphics{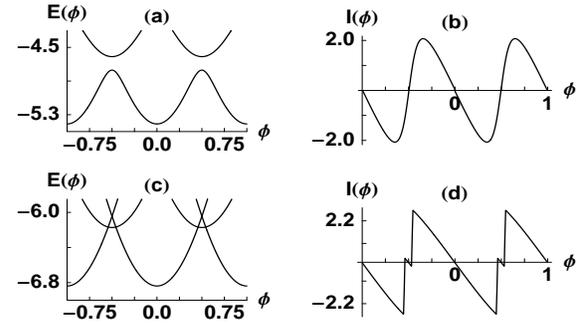}}\par}
\caption{The $E-\phi$ and $I-\phi$ curves for following six ($\uparrow,
\uparrow,\uparrow$, $\downarrow,\downarrow,\downarrow$) electron systems :
i) $N=6$, $U=2$ in (a) \& (b) and ii)
$N=7$, $U=2$ in (c) \& (d).}
\label{energy5}
\end{figure}

Lastly, we consider six electron systems, and
investigate the nature of persistent current in 
rings with three up and three down spin electrons. At half-filling 
(a benzene-like case with $N=6$ and $n=6$), we see that
$I(\phi)$ becomes a continuous function of $\phi$ as
shown in Fig.~\ref{energy5}(b), 
exactly similar to the half-filled four electron case.
Fig.~\ref{energy5}(c) and Fig.~\ref{energy5}(d) are respectively the plot of
$E-\phi$ and $I-\phi$ curves for a typical non-half-filled case with
$N=7$ and $n=6$, and,
we find striking similarity in the behavior of
persistent current with other non-half-filled systems containing even
number of electrons. Hence it becomes apparent that mesoscopic Hubbard
rings with even number of electrons exhibit similar characteristic
features in the persistent current. 

In the present context it is also interesting to study the conducting
properties of these mesoscopic rings in the presence of Hubbard correlation.
For this purpose, we calculate Drude weight $D$ of these systems using
Eq.~(\ref{drude}).  
A metallic phase (ballistic regime) will be characterized
by a finite value of $D$, while it will be zero in an insulating phase
(localized regime)~\cite{kohn}.
\begin{figure}[h]
{\centering \resizebox*{8.0cm}{5.5cm}{\includegraphics{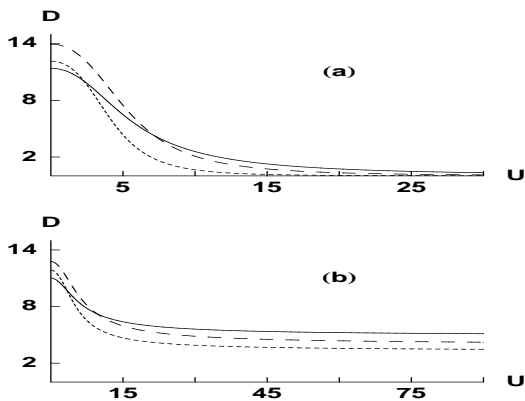}}\par}
\caption{Drude weight vs. interaction for (a) half-filled 
and (b) non-half-filled systems. 
The solid, dashed and dotted lines respectively corresponds to
 3, 4 and 5 electron cases.}
\label{energy6}
\end{figure}
In Fig.~\ref{energy6}(a) and
Fig.~\ref{energy6}(b), we plot $D$ as a function of 
$U$ respectively for half-filled and non-half-filled rings with 
$n=3, 4$ and $5$.  In the non-half-filled cases, the number of sites
corresponding to a given value of $n$ is taken as $N=n+1$, and, we
note that other choices of $N$ do not affect the basic features of 
the $D$ versus $U$ curves. It is evident from Fig.~\ref{energy6}(a) that
for low $U$, the half-filled systems are in the metallic phase,
and, they become insulating only when $U$ is very large. 
In the insulating phase ground state do not support any empty site,
and the situation is somewhat analogous to Mott localization in
$1$D infinite lattices. Fig.~\ref{energy6}(b) shows that the non-half-filled
rings are always conducting irrespective of the value of $U$. 

In conclusion, we studied exactly the persistent current and Drude
weight in small mesoscopic Hubbard rings at zero temperature, and,
identified that the Hubbard correlation leads to many interesting new
results. The main results are : the appearance of kinks in the 
persistent current, observation of both $\phi_{0}$ and $\phi_{0}/2$
periodic currents, no singular behavior of persistent current in 
the half-filled systems with even number of electrons,
evidence of $U$-independent eigenstates, existence of both
metallic and insulating phases, etc. We also observe discontinuities
in persistent current at non-integer values of $\phi_0$ due to 
electron-electron correlation, which crucially depends on the filling
of the ring and also on the parity of the number of electrons. This
gives rise to anomalous Aharonov-Bohm oscillations in the persistent 
current with much reduced period where periodicity is not perfect,
and Keyser {\em et al.} \cite{keyser} experimentally observed similar
anomalous Aharonov-Bohm oscillations in the conductance of a few-electron
quantum ring. Finally, it is worth noting that the above mentioned
effects of interaction will be pronounced in strongly
correlated electron systems, like, mesoscopic rings of transition
metal elements.

We thank R. K. Moitra
and B. Bhattacharya for many illuminating discussions.

\end{document}